\documentclass[journal]{IEEEtran}

\newcommand{\mycol}{1}
\usepackage{tikz}
\usepackage{mathtools}
\usepackage{amsmath}

\begin{document}

\title{Dual on-chip SQUID measurement protocol for flux detection in large magnetic fields}

\author{Josiah~Cochran,	Giovanni~Franco-Rivera, Denghui~Zhang, Lei~Chen, Zhen~Wang and~Irinel~Chiorescu%
\thanks{J. Cochran G. Franco-Rivera and I. Chiorescu are with the Department of Physics, Florida State University and the National High Magnetic Field Laboratory, 1800, East Paul Dirac Drive, 32310, Tallahassee, FL, USA. D. Zhang, L. Chen and Z. Wang are with the Center for Excellence in Superconducting Electronics, Shanghai Institute of Microsystem and Information Technology, Chinese Academy of Sciences, Shanghai 200050, China.}
\thanks{Manuscript received 2020}}

%



\maketitle

\begin{abstract}
Sensitive magnetometers that can operate in high magnetic fields are essential for detecting magnetic resonance signals originating from small ensembles of quantum spins. Such devices have potential applications in quantum technologies, in particular quantum computing. We present a novel experimental setup implementing a differential flux measurement using two DC-SQUID magnetometers. The differential measurement allows for cancellation of background flux signals while enhancing sample signal. The developed protocol uses pulsed readout which minimizes on-chip heating since sub-Kelvin temperatures are needed to preserve quantum spin coherence. Results of a proof of concept experiment are shown as well. 
\end{abstract}


%
\IEEEpeerreviewmaketitle

\section{Introduction}
%
%
%
\IEEEPARstart{S}{olid} state spin qubits are prime candidates for the next generation of quantum computers due to their long coherence times~\cite{bertaina2020experimental}. Spin qubits have been realized experimentally using molecular magnets~\cite{bertaina2008quantum,ardavan2007will,shiddiq2016enhancing}, diluted quantum spins in various matrices~\cite{bertaina2017forbidden}, and quantum dots~\cite{yang2020operation,moreno2018molecular}. Although single spin detection is possible with quantum dots, the transport methods used to detect the spins can be detrimental to the quantum state leading to short coherence times compared to magnetic detection which only couples weakly to the spins~\cite{godfrin2017electrical,morello2018scalable}. The weak magnetic coupling, although giving improved coherence times, has the downside of being difficult to implement. Classical Electron Spin Resonance experiments use 3-D microwave cavities tuned near resonance and a detection of increased microwave absorption indicates the presence of a spin system. The straight forward application of this to on-chip devices is the use of thin film resonators~\cite{groll2010measurement}. Recent attempts to improve magnetic spin sensitivity have included tunable bifurcation resonators~\cite{budoyo2018electron}, coupling to an artificial atom~\cite{toida2019electron}, coupling to a nano fabricated inductance~\cite{probst2017inductive}, and a single Direct-Current Superconducting QUantum Interference Device (DC-SQUID)~\cite{chen2016high,yue2017sensitive}. DC-SQUIDs are extremely sensitive flux detectors and provide much promise for sensitive spin detection. Niobium nano-bridge junctions are one of the strongest candidates being on the order of tens of nanometers and hundreds of atoms across therefore maintaining their mesoscopic properties compared to nanometer size carbon nano-tube junctions which function as a quantum dot~\cite{Cleuziou2006}. The mesoscopic size of Niobium nano-bridge junctions provides weak magnetic coupling to spin systems while still being sensitive enough to detect a microscopic spin sample~\cite{bouchiat2009detection}. 

Previous measurements involving a single DC-SQUID have been performed with Gd spin systems~\cite{yue2017sensitive}. Measurements involving multiple DC-SQUIDs in a gradiometer setup have been performed, but these methods are usually done with the SQUID isolated from the magnetic field in a shielded enclosure and coupled to the sample via a flux transformer~\cite{koch1993three,stolz2020long,savukov2020detection}. Direct coupling to spins placed in the planar magnetic field allows for a much more sensitive detection. This work presents a method using two DC-SQUIDs to perform a differential flux measurement in order to improve spin sensitivity and cancel out background signals. The reason that these DC-SQUIDs can be placed in field is that they are very thin (20~nm) therefore the cross sectional area experiencing the magnetic field is very small and superconductivity can last up to a few Tesla~\cite{chen2010chip}. The following sections will outline the setup and measurement procedure, the proof of concept results as well as a noise and sensitivity evaluation.

\section{SQUID Fabrication}
\begin{figure}[!t]
	\centering
	\includegraphics[width=\mycol\columnwidth]{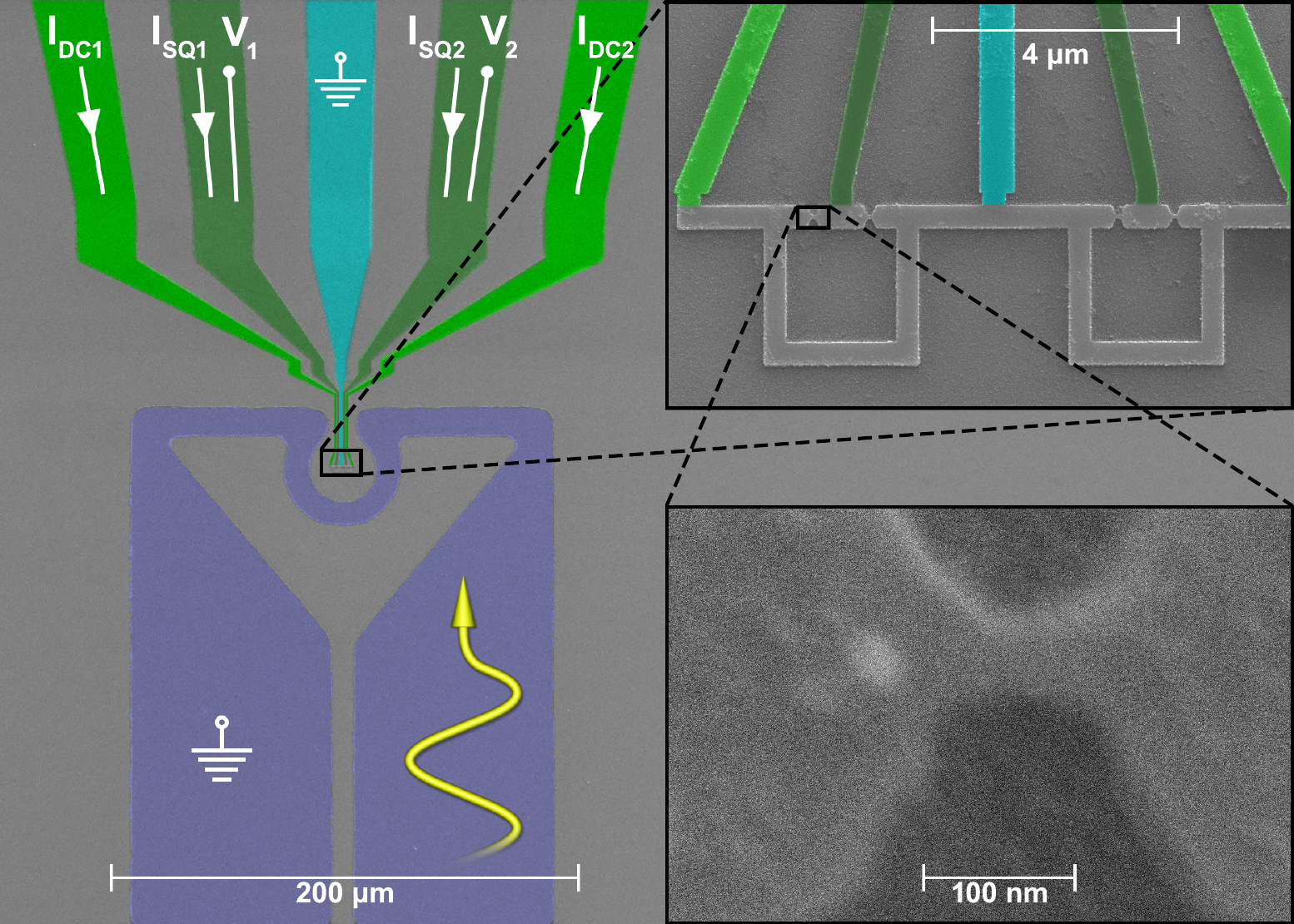}
	\caption{Scanning Electron Micrograph (SEM) images of the Nb microwave-SQUID device at different magnifications; the scales are indicated in white, the purple shows the Nb waveguide terminated in an $\Omega$-loop, the light green are flux bias lines for each SQUID, the dark green indicate the current bias and voltage pickups and the middle line is the ground. The two SQUIDs (top inset) have nanobridge junctions (bottom inset) on the upper branch with the bias current lines located in-between the junctions.}
	\label{SEM}
\end{figure}
 
The fabrication of the SQUIDs begins with a 20~nm thin film of Nb that has been deposited on top of a Si substrate by DC-magnetron sputtering. A 50~$\Omega$ microwave line is patterned in an optical lithography stage and an Al mask is deposited. A second optical stage is done to create tall alignment markers in Al that will be visible for the electron-beam writing. The wafer is spin coated in PMMA resist and the SQUID pattern is written using JC Nabity NPGS system connected to a Helios G4UC SEM. An Al mask for the SQUIDs is deposited and the chip is etched in a CF$_4$ environment using a reactive ion etcher. The Al is dissolved in NaOH revealing the final pattern. The resulting Nb chip is shown in Fig.~\ref{SEM}: the 50~$\Omega$ microwave co-planar strip line is terminated in an $\Omega$-loop which contains the two SQUIDS (top inset) featuring nano-bridge junctions (bottom inset). Two DC lines provide switching currents in-between the Josephson junctions (dark green) and two lateral lines are used for flux biasing each SQUID independently (light green). 

\section{Setup and Procedure}
\label{Setup}
\begin{figure}[!t]
	\centering
	\includegraphics[width=\mycol\columnwidth] {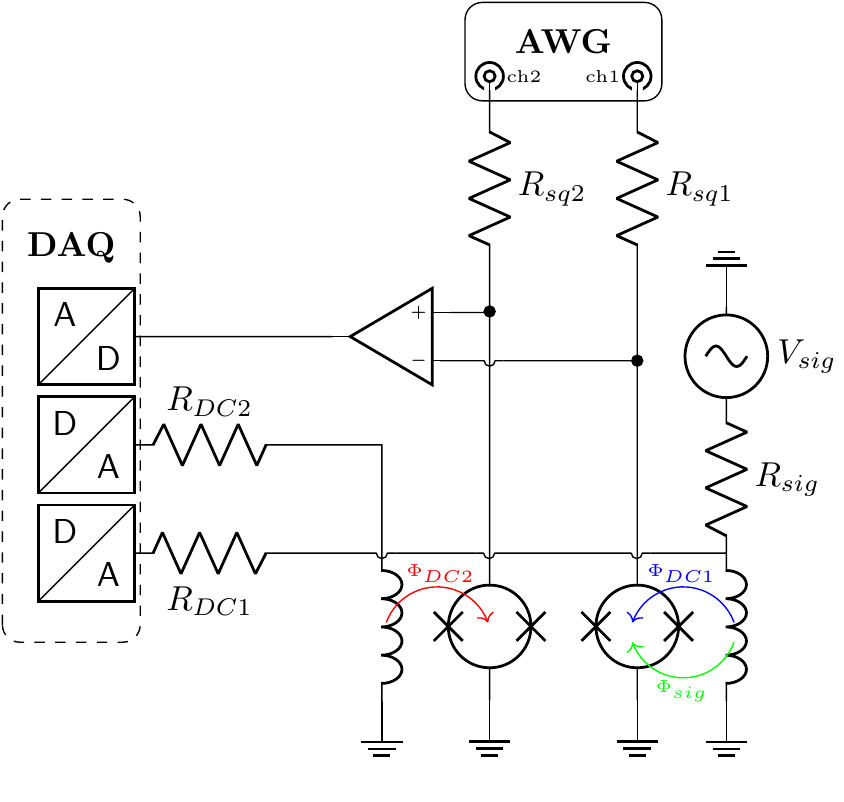}
	\caption{An AWG is used to generate time-shifted current pulses on the SQUIDs using inline resistors $R_{sq1,2}$. SQUID voltages can thus be individually be picked up by a single amplifier and sent to the analog to digital converter of a DAQ device. The DAQ uses two digital to analog outputs with inline resistors $R_{DC1,2}$ to flux bias the SQUIDs by creating fluxes $\Phi_{DC1,2}$. Thus, a feedback loop locks the total flux in each SQUID at a fixed value. $SQ_1$ is the sample SQUID while $SQ_2$ measures the background. A function generator $V_{sig}$ is used to simulate a sample signal $\Phi_{sig}$ for a proof of concept test.}
	\label{c3}
\end{figure} 

An overview of the electronics setup used in order to execute the dual SQUID measurement is shown in Fig.~\ref{c3}.  A two-channel arbitrary waveform generator (AWG) is used with series resistances $R_{sq1,2}$ in order to send current pulses to SQUIDs. A SRS911 pre-amplifier detects the difference between the voltages across SQUIDs; the current pulses are slightly shifted in time such that the amplified signal contains the signals of the individual SQUIDs as voltage pulses at different moments in time. An ADwin digitizer is used to readout this signal and to provide feedback currents to bias the flux of each SQUID. 

The duration of the current pulses is $\sim1-10$~$\mu$s such that the DC-characteristics of the SQUID are valid. A number of $N=$ 250 pulses are sent in order to calculate SQUIDs' switching probabilities $P_{sw1,2}$.  The repetition frequency is low enough to avoid heating as well as to allow quasi particles to relax after a switching event. The pulses height is gradually increased and the proportion of pulses leading to a finite SQUID voltage defines the switching probability. The switching current $I_{sw}$ is defined as the pulse height for which $P_{sw}=50\%$. The ADwin feedback algorithm is built to measure the amount of flux bias needed to keep the SQUIDs' switching probabilities at 50\%. The lines $DC_{1,2}$ in Fig.~\ref{c3} provide such feedback information since they bias SQUIDs$_{1,2}$ with bias fluxes $\Phi_{DC1,2}$, respectively. For testing purposes, the flux of a sample is simulated using a function generator $V_{sig}/R_{sig}$ on the same line as the flux bias of SQUID$_1$. The device is mounted on a sample holder attached to the mixing chamber of a dilution refrigerator and placed inside a superconducting magnet. The magnet is used to simulate a background static magnetic flux, for the tests presented in this work.

\begin{figure}[!t]
	\centering
	\includegraphics[width=\mycol\columnwidth]{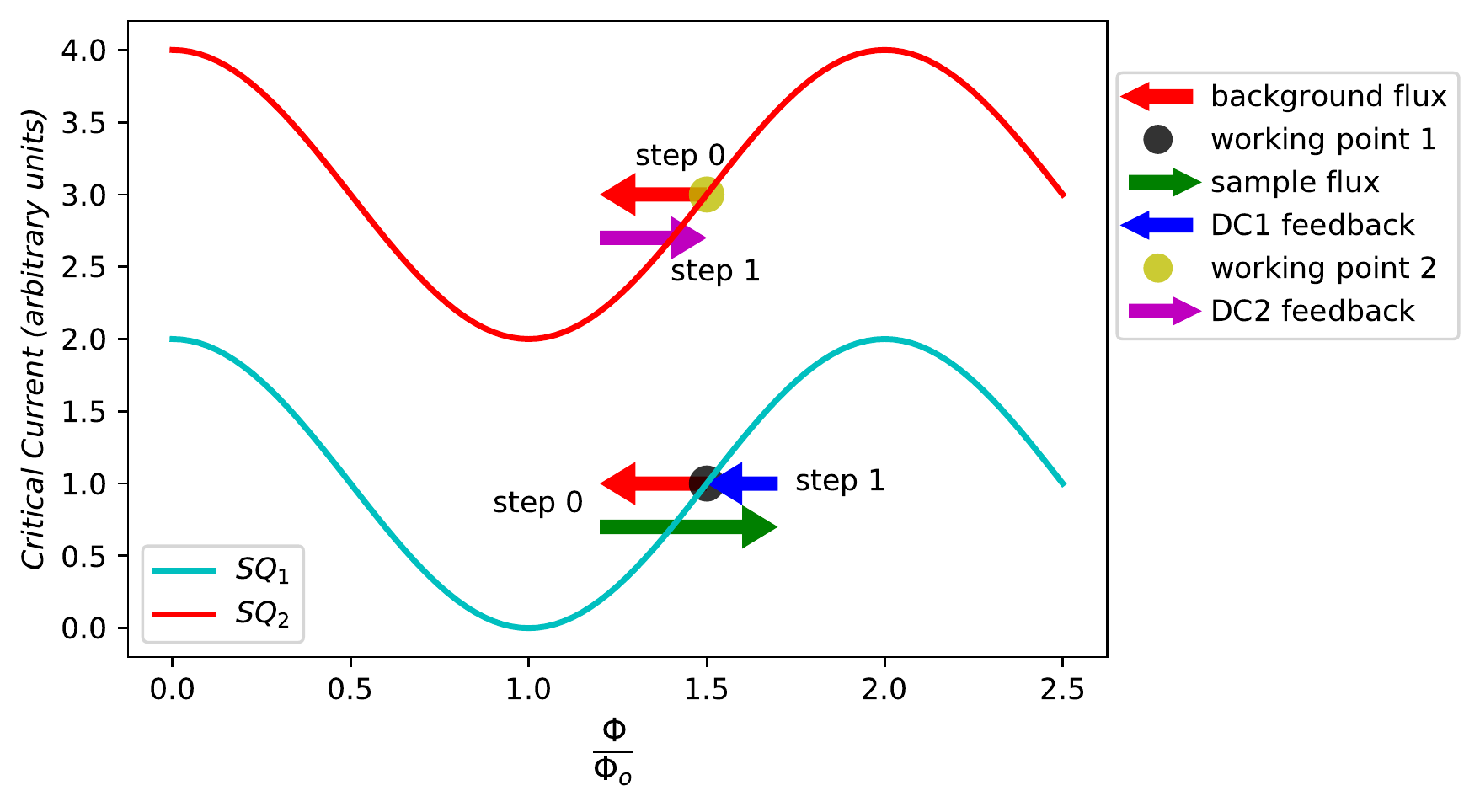}
	\caption{Sketch showing the SQUIDs switching currents as a function of flux and the implemented measurement protocol. Step 0: $SQ_2$ is shifted by the background flux while $SQ_1$ is shifted by background plus sample flux. Step 1: both feedback currents find their respective working points. $I_{DC2}$ has to be compensated from $I_{DC1}$ after the fact to obtain the sample flux.}
	\label{m3}
\end{figure} 
The sequence of events in the feedback algorithm used in order to enact the differential measurement is presented in Fig.~\ref{m3}. First the $I_{sw1,2}$ modulation curves must be obtained for each SQUID as a function of $\Phi_{DC1,2}$ as sketched in Fig.~\ref{m3}: a typical modulation shows a $\Phi_0$ periodic switching current. Indeed, both $I_{sw}$ and $P_{sw}$ change periodically with the magnetic flux penetrating the SQUID generated by $\Phi_{DC1,2}$ and/or the external field of a superconducting coil common to both SQUIDs. Next, a working point is chosen on each modulation curve with a constant current pulse height and initial DC flux bias in order to obtain a  $P_{sw}\sim$ 50\%. $SQ_2$ is pulled off of the working point by a background flux while $SQ_1$ is pulled off its working point by the superposition of the background flux (red arrow) and the sample flux (green flux). The feedback algorithm begins to search with the flux bias lines for $P_{sw}=$ 50\% (blue and purple arrows). The change in the current in each DC bias line is proportional to the flux seen by each SQUID. The difference between feedback currents on $DC_{1,2}$ is proportional to the sample signal with any background fluctuations removed. 

In practice however, the periodicity in $\Phi_{DC1,2}$ is slightly different between the two SQUIDs due to a very small difference in parameters such as the SQUIDs areas and the DC bias coupling. For that reason, the initial calibration procedure aims at finding a factor (very close to unity) to multiply $I_{DC2}$ such that it compensates $I_{DC1}$ for $V_{sig}=0$ but in variable external field. This factor is then always to be multiplied with $I_{DC2}$, for the rest of the experiments.

\section{Proof of Concept Results}

\begin{figure}[!t]
	\centering
	\includegraphics[width=\mycol\columnwidth]{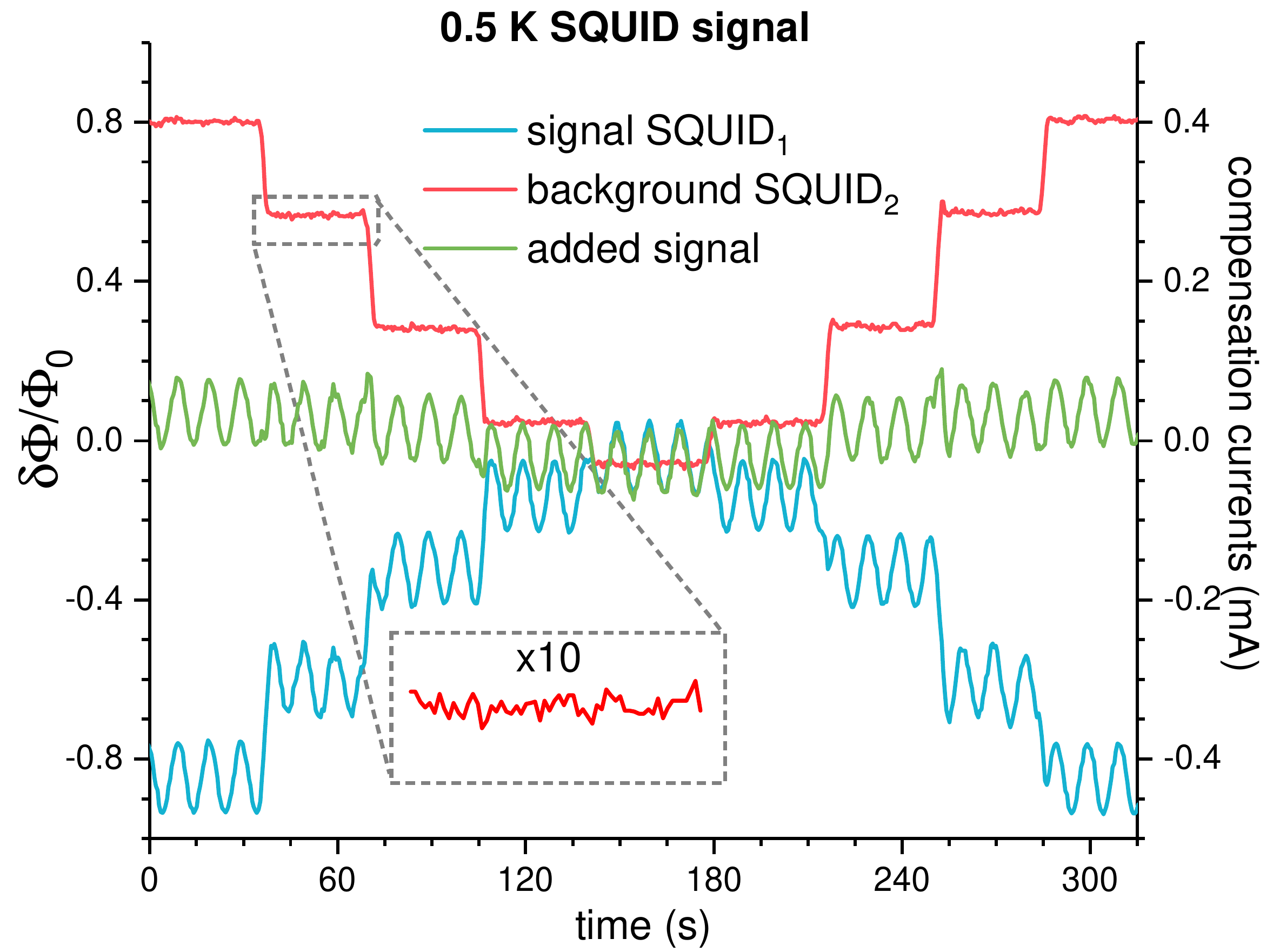}
	\caption{Experimental results at 0.5~K. The blue and red curves shows the $I_{DC1,2}$ bias currents in mA (right axis) and $\Phi_0$ units (left axis), respectively. The green curve is the difference of their absolute values taking into account the calibration factor between the SQUIDs (see text) which is a measure of the simulated sample flux. The insert is a $\times10$ magnification of one of the nine plateaus of $I_{DC2}$, used to calculate the noise level (see text).}
	\label{0.5k}
\end{figure}
Results from a measurement taken with the SQUIDs at 500~mK are given in Fig.~\ref{0.5k} ; the temperature is constantly monitored with a thermometer placed in the sample's vicinity (no heating effects were noted during the SQUID operation). The sample, fabricated at SIMIT, has the functionalities described in Fig.~\ref{SEM}. Before running the experiment, the calibration procedure discussed above is performed. Then, a sinusoidal signal $V_{sig}$ is applied onto the feedback line of $SQ_1$ from the function generator to create a sinusoidal sample flux and the vector magnet is stepped in small increments in order to simulate a changing background flux. The left axis of Fig.~\ref{0.5k} indicates detected changes in $\Phi_0$ units. The effect of the external coil is detected by the background SQUID, here called SQUID$_{2}$ (red line).

The blue curve shows the flux measurement of $SQ_1$. As presented above, the feedback mechanism implemented with the aide of the DAQ holds the SQUIDs at $P_{sw1,2}=50\%$ while providing a fast measurement of the required $I_{DC1,2}$. In the current test, the repetition time of the $N=250$ pulses was 3.3~kHz and the search algorithm was able to converge to $P_{sw1,2}=50\%$ within few points, leading to a timing of about 0.5~s per data point.

Once the search is done, the calibrated difference of the absolute values of $I_{DC1,2}$ gives the green curve which is proportional to the sample flux. The proportionality is ensured by the linear dependence of $I_{sw}$ vs flux in the immediate vicinity of the search area (in first order). Additionally, the speed of the search algorithm ensures that the flux-induced changes in the switching current are small and thus the recovery back to the working point of each SQUID is done efficiently.  

The recovered sample signal (in green) shows the expected sinusoidal behavior, although a small deviation from a flat line can be observed for the baseline of the signal. This can be attributed to small changes in the kinetic inductance of the Nb devices, and thus in the coupling of the bias lines, in the presence of an external field. The effect can be easily reduced by further calibration procedures. 

An important aspect is the level of noise affecting the detected $I_{DC1,2}$. An analysis of the flat plateaus of $I_{DC2}$ (red curve) indicate an RMS flux noise of 5.32~m$\Phi_0$, generated mostly by the AWG and which will be further analyzed below. The $P_{sw}$ feedback and convergence mechanism can also generate small oscillations picked up within the RMS value.

\section{Noise and sensitivity analysis}
\begin{figure}[!t]
	\centering
	\includegraphics[width=\mycol\columnwidth]{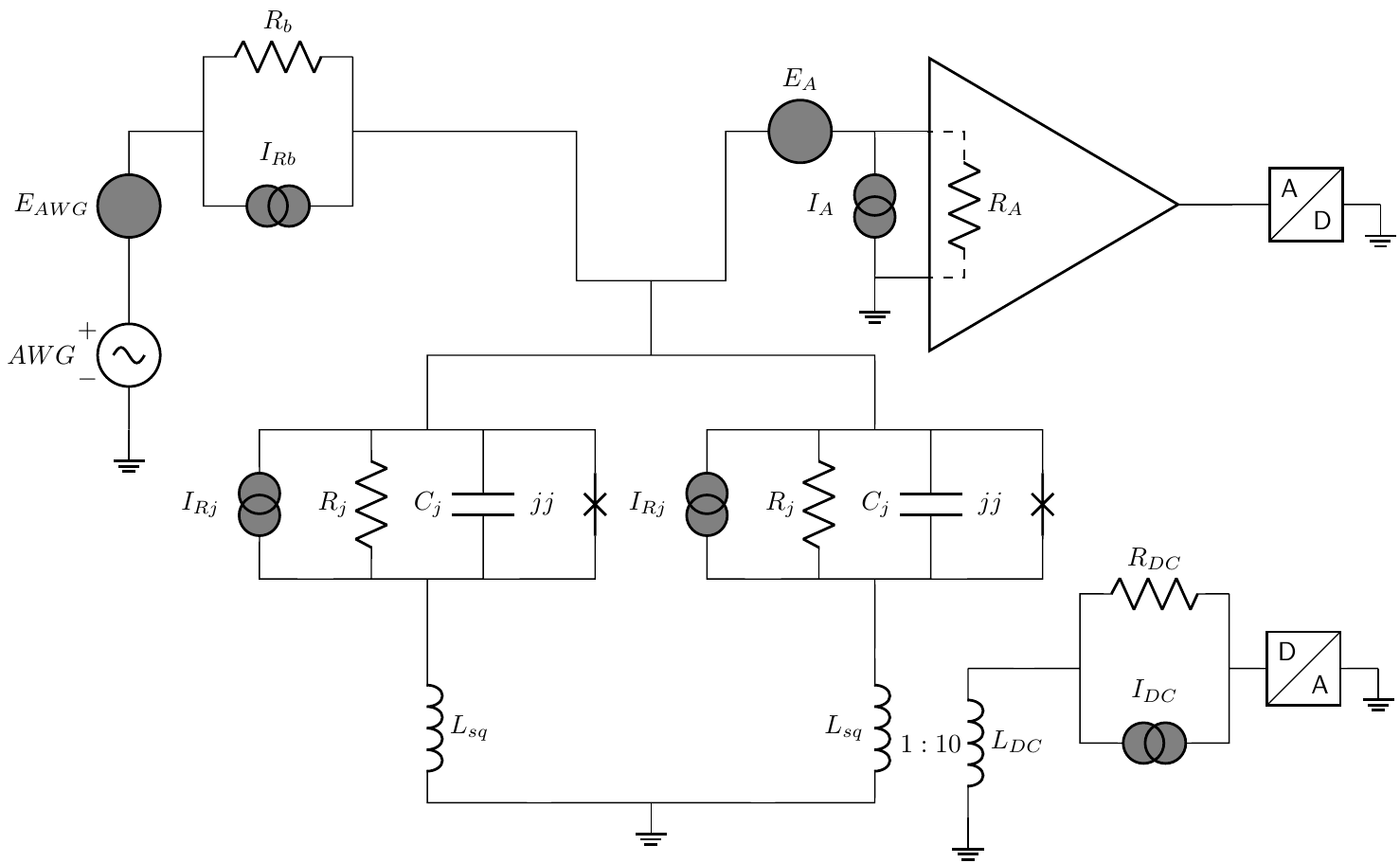}
	\caption{Effective noise model for electronics setup. Gray voltage and current sources are associated component noise.}
	\label{noise_ckt}
\end{figure}  

\begin{table}[]
    \centering
    \caption{RMS Noise from each source}
    \label{noise_table}
    \begin{tabular}{|c|c|c|}
    \hline
    Source  &  noise (nA) & M\\
    \hline
    \hline
    $E_{AWG}/R_b$   &  277.4 & 10\\
    \hline
    $I_{Rb}$    &  7.4  & 10\\
    \hline
    $I_{A}$    &   4.8 & 10\\
    \hline
    $E_{A}/R_A$    &  72 & 10\\
    \hline
    $I_{Rj}$    &  8.1 & 10\\
    \hline
    $I_{DC}$    &  16 & 1\\
    \hline
    \hline
    Total $\sqrt{\sum (M\delta)^2}$ &   2870 & - \\
    \hline
    Measured & 2660 & - \\
    \hline
\end{tabular}
\end{table}

The spin sensitivity of the setup can be determined by looking at the noise level of the detected signal, namely the bias currents $I_{DC1,2}$. The horizontal plateaus visible in $I_{DC2}$ (red curve in Fig.~\ref{0.5k}) are isolated for data analysis and the standard deviation of each plateau is calculated; one such plateau is shown in the insert of the figure. The average of all nine values is obtained to be $\delta\Phi/\sqrt{N}=5.32$~m$\Phi_0$ or $\delta I_{DC2}/\sqrt{N}=2.66$~$\mu$A, to be compared against the RMS noise, which is assumed Gaussian, generated by the electronic setup.

An effective noise model of the electronic setup is presented in Fig.~\ref{noise_ckt} for a single SQUID (see \cite{motchenbacher1993} regarding noise circuit modeling). A gray circle represents a noise source for the circuit element next to it: $E_{AWG,A}$ for the AWG and the voltage amplifier respectively, $I_{Rb,DC}$ for the current and flux bias respectively and $I_{Rj}$ the SQUID junction noise due to a potential small normal state resistance. The bias resistances are $R_b=4.7$~k$\Omega$ and $R_{DC}$=1~k$\Omega$. The value of $R_j=$6.8~$\Omega$ at 0.5~K is estimated as the normal state resistance immediately after switching \cite{granata2011noise}. The Johnson-Nyquist noise current for each of the resistances given above is:
 
\begin{equation}
	\label{JNN}
	\delta I = \sqrt{4k_BTB_w/R}
\end{equation}
where $k_B$ is the Boltzmann factor, $T=290$~K for $R=R_{b,DC}$ and 0.5~K for $R=R_j$ while $B_w=16$~MHz is the measured bandwidth of the setup wiring. 

The back-action noise of the amplifier on the SQUID is modeled with a noise free amplifier, a noise free resistance $R_A=100$~$\Omega$ across the amplifier inputs and two uncorrelated noise sources $I_A$ and $E_A$, for input noise current and voltage, respectively~\cite{motchenbacher1993}. The values for the amplifier noise model are taken directly from the SIM911 amplifier datasheet.

The power spectral density of the noise output of the AWG is measured with a HP3506A dynamic signal analyzer and divided into two ranges: low-frequency where the noise is $1/f$-like from 0 to the corner frequency $B_0=13$~kHz and white (flat) noise from $B_0$ to $B_w$. Therefore, the total noise voltage $E_{AWG}$ is given by:
\begin{equation}
	\label{Nawg}
	E_{AWG} = \sqrt{\int_{0}^{B_0} E_{1/f}^2(f) \,df} + E_{flat}\sqrt{B_w-B_0} = 1304 \mu V
\end{equation}
with $E_{1/f}(f) = \frac{4}{f+3400}$~mV/$\sqrt{\text{Hz}}$, and  $E_{flat} = 311$~nV/$\sqrt{\text{Hz}}$. The value $E_{AWG}/R_b$ gives the noise current generated by the AWG and it is the leading noise source in our current electronic setup.

The noise current from the AWG, the amplifier and potentially from the normal current component in the SQUID junctions, is leading to fluctuations in the SQUID switching current $I_{sw}$. As shown in Fig.~\ref{m3}, the derivative $\delta I_{sw}/\delta I_{DC}$ at the working point, transforms the noise of $I_{sw}$ into noise in the detected $I_{DC}$. For both SQUIDs, the slope at the working point is measured to be $\approx50$~$\mu$A$/\Phi_0$, while the modulation period as a function of $I_{DC}$ is $\approx$ 0.5~mA$/\Phi_0$. This results in a 10:1 coupling of $I_{DC}$:$I_{sq}$ as shown in Fig.~\ref{noise_ckt} using two coupled inductors. Therefore, the corresponding noise currents are to be multiplied by a factor of $M=10$ to evaluate the noise detected on the DC bias line.

The numerical values are given in Table~\ref{noise_table}. The first column lists the noise sources as presented above. The second column gives the numerical value of the corresponding current noise in nA. The third column indicates the coupling factor to the DC bias line which is $M=10$ for all sources except the DC bias source itself ($M=1$). The total contribution is given by the square root of the sum of all variances multiplied by their coupling factor and adds up to a total $\delta I_{DC}=2.87$~$\mu$A. The noise model presented here agrees very well with the measured RMS of 2.66~$\mu$A.

It can be noted that the main noise contribution is by far the one generated by the arbitrary waveform generator. Using a different generator, possibly on battery supply and increase $R_b$ by two orders of magnitude may alleviate this issue. At room temperature, the preamplifier and to a lesser extent, the bias resistances $R_{DC1,2}$ are relevant as well. These components can be made operable at low temperature and be inserted inside the refrigerator, at 4~K thus reducing their influence by about two orders of magnitude \cite{ivanov2011cryogenic}. The bandwidth of the circuit lines allows for fast pulsing of the SQUIDs but an effective operation could be achieved with a tenth of $B_w$ thus further reducing the noise. Overall, we estimate that two orders of magnitude improvement could be achieved in the future. If the electronics setup is optimized for $\mu\Phi_0$ sensitivity, theoretical estimations predict that a number of hundreds of Bohr magnetons are detectable~\cite{bouchiat2009detection}. With the setup as presented here, this number is larger by a factor of $\sim10^4$.

\section{Sample Layout}

\begin{figure}[!t]
	\centering
	\includegraphics[width=\mycol\columnwidth]{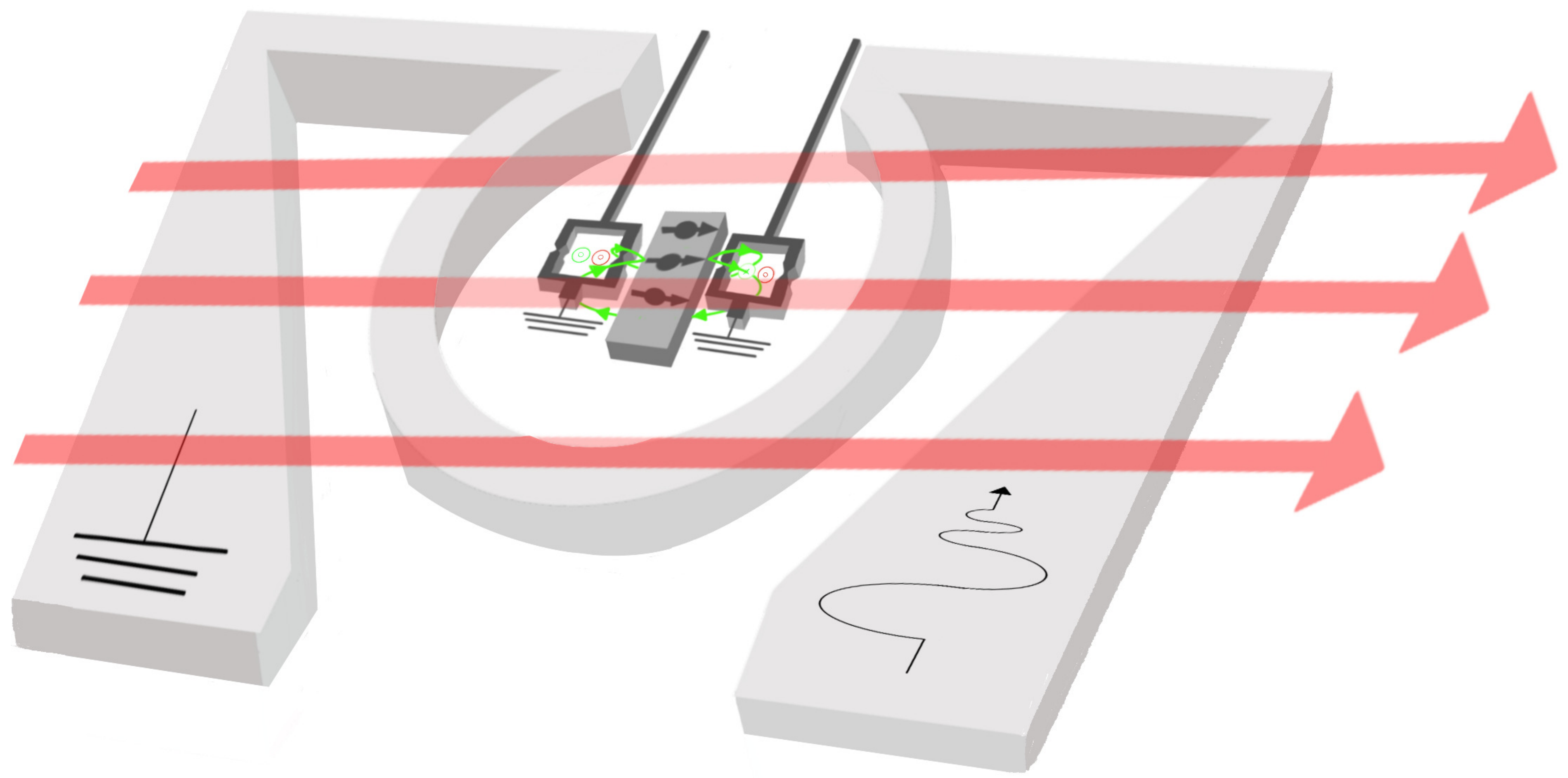}
	\caption{A spin system (middle sample) in magnetic field (red arrows) is placed between two SQUIDs so that both SQUIDs measure the spin with opposing polarity $\Phi_{SQUID1}=\Phi_{spin}(t)+\Phi_b(t)$~and~ $\Phi_{SQUID2}=-\Phi_{spin}(t)+\Phi_b(t)$. The green lines represent the flux lines of the magnetic sample. The background is canceled and sensitivity is double upon taking the difference $\Phi_{SQUID1}-\Phi_{SQUID2}=2\Phi_{spin}(t)$. Furthermore the noise is reduced as well. Graphic produced by Nayt Cochran~\protect\footnotemark[1].}
	\label{diff_squid}
\end{figure}  
\footnotetext[1]{@cartoonographer , Nashville, TN USA}

The device discussed here allows highly sensitive detection of spins while, at the same time, implementing microwave elements needed to control quantum spins. We note that a pulsed microwave operation does not impede the SQUID readout as long as the delay between them can be optimized to capture the spin information. We now present a sample-SQUID configuration which can enhance the signal while reducing the noise by adjusting the SQUIDs separation to sample size. The proof of concept test (Fig.~\ref{0.5k}) assumes a layout where one SQUID is far away from the sample and one is very close. It is however possible to micro-manipulate a small magnetic sample on a chip and position it directly between the two SQUIDs as shown in Fig.~\ref{diff_squid}. 

Assuming a strong external magnetic field (red arrows), the flux lines of the sample (in green) are penetrating the two SQUIDs in an opposite fashion. Thus, the differential readout of the SQUIDs will cancel out the background fluctuations while doubling the signal:
\begin{equation}
\label{eq_spin}
\Phi_{SQUID1} - \Phi_{SQUID2}=2\Phi_{spin}
\end{equation}
and 
\begin{equation}
\label{eq3}
\sigma_{spin}^2 =  \frac{1}{4}\sigma_{SQUID1}^2 + \frac{1}{4}\sigma_{SQUID2}^2 = \frac{1}{2}\sigma_{SQUID}^2 \\
\end{equation}
where it is assumed that the two SQUIDs have similar RMS figures $\sigma_{SQUID}=\sigma_{SQUID1,2}$.  Therefore $\sigma_{spin} = \frac{1}{\sqrt{2}}\sigma_{SQUID}$ leading to an increase in spin sensitivity by $\sqrt{2}$. In practice, the two SQUIDs will have slightly different circuit parameters leading to an imperfect compensation in Eq.~\ref{eq_spin}. However, this effect can be calibrated out in the absence of a sample, in a similar way as described in Sect.~\ref{Setup}.

Note that the proof of concept measurements shown in this paper presents a method applied to a sample that couples to only one SQUID while a second SQUID is used to remove any background signals. The factor of two in Eq.~\ref{eq_spin} drops and the overall electronic noise is increased by $\sqrt{2}$ in this case.

\section{Conclusion}
This work presents a novel spin detection protocol based on a sensitive differential detection of magnetic signals, which can impact future quantum technologies. Proof of concept experimental results for this method are shown together with a noise and sensitivity analysis. Future development of the protocol is described as well. The presented method shows much promise due to the sensitivity of DC-SQUIDs combined with the differential topology that allows the SQUIDs to be directly coupled to the sample while placed in a large planar magnetic field.

\section*{Acknowledgment}

J.C., G.F.-R. and I.C. thank the support from the National Science Foundation Cooperative Agreement No. DMR-1644779 and the State of Florida. D.Z., L.C. and Z.W. acknowledge the support from the Frontier Science Key program and the Strategic Priority Research program of the CAS (Grant No. QYZDY-SSW-JSC033, XDA18000000) and the National Science Foundation of China (Grant No. 11827805, 62071458).

\bibliographystyle{IEEEtran}
\bibliography{Diff_SQUID_TAS2021}

\end{document}